\pdfoutput=1
\documentclass{jpp}

\usepackage[utf8]{inputenc}
\usepackage[T1]{fontenc}
\usepackage[hidelinks]{hyperref}
\usepackage{amsmath, bm, cancel, graphicx, tikz, ulem, xcolor}
\definecolor{myRed}{rgb}{.84,0,.08}
 
\usetikzlibrary{patterns}

\title{Generalized Expanding-Box Formulations of Reduced Magnetohydrodynamics in the Solar Wind}

\shorttitle{Expanding-Box Formulations of RMHD}
\shortauthor{V. David et al.}

\author{
V. David \aff{1} \corresp{\email{vincent.david@unh.edu}},
B. D. G. Chandran \aff{1},
R. Meyrand \aff{1,2},
J. Squire \aff{2},
E. L. Yerger \aff{1}
}

\affiliation{
\aff{1}Space Science Center and Department of Physics and Astronomy, University of New
Hampshire, Durham, NH 03824
\aff{2}Physics Department, University of Otago, Dunedin 9010, New Zealand
}

\begin{document}
\maketitle

\begin{abstract}
We derive a set of simplified equations that can be used for numerical studies of reduced magnetohydrodynamic turbulence within a small patch of the radially expanding solar wind. We allow the box to be either stationary in the Sun's frame or to be moving at an arbitrary velocity along the background magnetic field lines, which we take to be approximately radial. We focus in particular on the case in which the box moves at the same speed as outward-propagating Alfvén waves. To aid in the design and optimization of future numerical simulations, we express the equations in terms of scalar potentials and Clebsch coordinates. The equations we derive will be particularly useful for conducting high-resolution numerical simulations of reflection-driven magnetohydrodynamic turbulence in the solar wind, and may also be useful for studying turbulence within other astrophysical outflows.
\end{abstract}

%----------------------------------------------
% NEW SECTION
%----------------------------------------------
\section{Introduction} \label{sec:Introduction}

Alfvénic turbulence consists of non-compressive fluctuations in the velocity $\boldsymbol{v}$ and magnetic field $\boldsymbol{B}$ that have comparable energy densities and is believed to play a key role in generating the solar wind.
The Sun launches Alfvén waves that travel outward and undergo partial reflection due to the radial variation in the Alfvén speed $v_{\rm A}$.
The reflected waves then interact nonlinearly with the waves still traveling outward, causing fluctuation energy to cascade from large scales to small scales, where the fluctuations dissipate, heating and accelerating the plasma.
Understanding the details of solar-wind turbulence has several important implications.
First, it helps us predict how much mass the Sun loses through the solar wind and the speed at which this wind travels.
The location where this turbulence heats the plasma is critical: heating inside the sonic critical point increases the mass loss rate, while heating farther away increases the wind’s final speed \citep{Leer_1980_JGR}.
Therefore, to determine the speed and mass flux of the solar wind, we need to understand how quickly this turbulence dissipates.
Additionally, the inertial-range power spectrum of the turbulence, which determines the strength and anisotropy of small-scale fluctuations, is important to understand.
These fluctuations are crucial because they control the amount of heating through mechanisms like stochastic heating and cyclotron heating \citep[see, e.g.,][]{Chandran_2010_ApJ, Squire_2022_NatAs}.
Finally, the degree of anisotropy in these fluctuations directly affects the efficiency of different dissipation mechanisms, which can control, for instance, the relative heating rate of different species \citep{Johnston_2024_arXiv}.

One of the most effective tools we have to study this turbulence is direct numerical simulations.
A widely used approach involves flux-tube simulations, where we model a narrow magnetic flux tube centered on a radial magnetic-field line and track the turbulence as it evolves along the tube \citep{Dmitruk_2003_ApJ, van_Ballegooijen_2011_ApJ, Perez_2013_ApJ, van_Ballegooijen_2016_ApJ, van_Ballegooijen_2017_ApJ, Chandran_2019_JPP}.
This method has shown great promise and has been extensively employed, but it presents a significant challenge: resolving the radial direction requires a vast number of grid points, leading to extremely high computational demands.
A potential remedy is the so-called ``expending-box model''.
The model simplifies the problem by focusing on a small, expanding parcel of plasma as it moves away from the Sun \citep{Grappin_1993_PRL, Dong_2014_ApJ, Tenerani_2017_ApJ, Montagud-Camps_2018_ApJ, Shi_2020_ApJ, Squire_2020_ApJ, Johnston_2022_PoP, Grappin_2022_ApJ, Meyrand_2024_JPP}.
The approach reduces computational demands while still capturing many essential characteristics of the turbulence.
By concentrating on this localized region, the expanding-box model allows one to track how turbulence evolves as the plasma journeys through space, without resolving the full extent of the flux tube.

In this paper, we advance the expanding-box model by developing a generalized version tailored for reduced magnetohydrodynamics (RMHD).
This new formulation allows the simulation box to either remain stationary in the Sun’s frame or move at any speed in the radial direction.
Additionally, we have reformulated the equations using scalar potentials and Clebsch coordinates.
This approach not only facilitates working with the flux-tube geometry but also enhances our understanding of the system while preserving the leading-order dynamics of the turbulent cascade.
We believe that two types of simulations will be particularly valuable.
First, the wave-frame box simulation, which follows the outward-propagating waves ($z^+$) that carry most of the energy generalizes the standard expanding-box model to sub-Alfvénic regions close to the Sun, provides a complementary perspective on the development of turbulence.
Second, the Eulerian box simulation at a fixed location in space will allow us to explore how turbulence evolves in a steady environment \citep{Siggia_1978_JFM, Passot_2022_JPP}.

The paper is structured as follows.
In Section \ref{sec:governing-equations}, we present the mathematical foundation of our generalized expanding-box model. Section \ref{sec:Scalar} details the scalar formulations of this model. Section \ref{sec:Linear} explores the linear dynamics, highlighting the significant effects of reflections compared to homogeneous RMHD.
Finally, Section \ref{sec:Conclusion} addresses the implications of our findings for understanding solar-wind turbulence and outlines potential directions for future research.

%------------------------------------------------------------
% NEW SECTION
%------------------------------------------------------------
\section{Governing equations} \label{sec:governing-equations}

In the solar wind, the energetically dominant fluctuations are elongated along the magnetic field, to a degree that increases towards smaller scales \citep{Horbury_2008_PRL, Sahraoui_2010_PRL, Chen_2012_ApJ}.
In the inertial range, in which the fluctuating magnetic field~$\delta \boldsymbol{B}$ is small compared to the root-mean-square (r.m.s.) magnetic-field strength~$B_{\rm rms}$, such anisotropic fluctuations are described by a version of RMHD \citep{Kadomstev_1974_SJETP, Strauss_1976_PoF, Schekochihin_2009_ApJSS} that is modified to account for the outflow and inhomogeneity of the solar wind  \citep[e.g.][]{Heinemann_1980_JGR, Verdini_2007_ApJ, Chandran_2009_ApJ, van_Ballegooijen_2011_ApJ, Perez_2013_ApJ, van_Ballegooijen_2016_ApJ}. We refer to this set of equations as inhomogeneous RMHD, or IRMHD.
In this section, we review the derivation of the IRMHD equations for solar-wind outflow within a narrow magnetic flux tube.

%------------------------------------------
\subsection{Narrow-flux-tube approximation}

To properly describe an expanding plasma, we first need to design a magnetic flux tube. We assume it to be aligned along a radial magnetic field line. Using spherical coordinates $(r, \theta, \varphi)$ where $r$ is the radial distance, $\theta$ is the polar angle, and $\varphi$ is the azimuthal angle, we take $\theta=0$ to coincide with the magnetic-field line upon which the flux tube centers and define
\begin{equation}
\bar{B}_0 \equiv B_0 (r, \theta = 0).
\end{equation}
We assume the background magnetic field $B_0$ is independent of $\varphi$ and has no $\varphi$ component:
\begin{equation}
    B_{0 \varphi} = \frac{\partial B_{0 r}}{\partial \varphi} = \frac{\partial B_{0 \theta}}{\partial \varphi} = 0,
    \label{eq:Bphi0}
\end{equation}
and we require that $\boldsymbol{B}_0$ is smooth around $\theta=0$ (i.e., $\nabla^2 \boldsymbol{B}_0$ remains finite), which implies that
\begin{equation} \label{eq:B0-finite}
    \left.\frac{\partial B_{0 r}}{\partial \theta}\right\vert_{\theta=0} = 0, \quad \text{and} \quad B_{0\theta}\Big\vert_{\theta=0} = 0.
\end{equation}
This results in $B_{0 r} = \bar{B}_0 (r)+ O(\theta)^2$. 
Given~(\ref{eq:Bphi0}), the condition~$\boldsymbol{\nabla} \cdot \boldsymbol{B} = 0$ can be written as
\begin{equation}  \label{eq:B0-div-free}
    \frac{1}{r^2} \frac{\partial}{\partial r} \left( r^2 B_r\right) + \frac{1}{r \sin \theta} \frac{\partial}{\partial \theta} \left( B_\theta \sin \theta \right) = 0.
\end{equation}
For a narrow flux tube,  $\theta \ll 1$, and the polar component of the magnetic field can be expanded as
\begin{equation}
\label{eq:B0theta_exp}
    B_{0 \theta} = \sum_{n=1}^{\infty} c_n \theta^n,
\end{equation}
where $c_n$ are coefficients we need to determine, and the sum starts at $n=1$ to satisfy Eq. (\ref{eq:B0-finite}). 
Upon substituting~(\ref{eq:B0theta_exp}) into~(\ref{eq:B0-div-free}), we obtain
\begin{equation}
    \frac{2 c_1}{r} + \frac{3 c_2 \theta}{r}= - \frac{1}{r^2} \frac{\partial}{\partial r} \left( r^2 \bar{B}_0 \right) + O(\theta)^2.
\end{equation}
From this, we find that
\begin{equation}
    c_1 = - \frac{1}{r^2} \frac{\partial}{\partial r} \left( r^2 \bar{B}_0 \right), \quad \text{and} \quad c_2 = 0,
\end{equation}
indicating that the polar component of the magnetic field satisfies:
\begin{equation} \label{eq:B0_theta}
    B_{0 \theta} = - \frac{\theta}{2 r} \frac{\mathrm{d}}{\mathrm{d} r} \left( r^2 \bar{B}_0 \right) + O(\theta)^3.
\end{equation}
To summarize, we assume a background magnetic field $\boldsymbol{B}_0 = \bar{B}_0 (r) \hat{\boldsymbol{e}}_r + B_{0\theta} \hat{\boldsymbol{e}}_\theta$, with $B_{0\theta}$ given by Eq. (\ref{eq:B0_theta}).
This represents the narrow-flux-tube approximation, which has been extensively used to study waves and instabilities in flux tubes (see, e.g., \citealt{Defouw_1976_ApJ, Roberts_1978_SolPhys, Spruit_1981_AA, van_Ballegooijen_2011_ApJ}).
Following the work of \cite{Perez_2013_ApJ}, we also assume that the flux tube has a non-constant square cross-section and use Cartesian coordinates $x$ and $y$, which are perpendicular to the central field line, such that $\sqrt{x^2 + y^2} \ll r$.
This allows us to focus on the vicinity of the central field line.
Finally, we neglect the Parker spiral effect, focusing our study on the inner heliosphere.

%------------------------------------------------------
\subsection{Inhomogeneous Reduced Magnetohydrodynamics}
Our starting point is the ideal compressible MHD equations:
\begin{subequations} \label{eq:compressible-MHD}
\begin{align}
 \label{eq:continuity}
    \frac{\partial \rho}{\partial t} &= - \boldsymbol{\nabla} \cdot \left( \rho \boldsymbol{u} \right), \\
 \label{eq:momentum}
    \rho \left( \frac{\partial \boldsymbol{u}}{\partial t} + \boldsymbol{u} \cdot \boldsymbol{\nabla} \boldsymbol{u}\right) &= - \boldsymbol{\nabla} p_\mathrm{tot} + \frac{\boldsymbol{B} \cdot \boldsymbol{\nabla} \boldsymbol{B}}{4 \pi}, \\
\label{eq:induction}
    \frac{\partial \boldsymbol{B}}{\partial t} &= \boldsymbol{\nabla} \times \left( \boldsymbol{u} \times \boldsymbol{B} \right),
\end{align}
\end{subequations}
where $\rho$ is the mass density, $\boldsymbol{u}$ is the velocity, $\boldsymbol{B}$ is the magnetic field, and $p_\mathrm{tot} = P + B^2/8\pi$ is the the sum of the thermal and magnetic pressures, respectively.
We set
\begin{subequations}
\begin{align}
    \boldsymbol{u}(\boldsymbol{r},t) &= U(\boldsymbol{r}) \hat{\boldsymbol{b}}_0 + \delta \boldsymbol{u}(\boldsymbol{r},t), \\
    \boldsymbol{B}(\boldsymbol{r},t) &= B_0(\boldsymbol{r}) \hat{\boldsymbol{b}}_0 + \delta \boldsymbol{B}(\boldsymbol{r},t), \\
     \rho(\boldsymbol{r},t) &= \rho_0(\boldsymbol{r}),
\end{align}
\end{subequations}
where the background flow velocity $\boldsymbol{U}$ is aligned with the background magnetic field $\boldsymbol{B}_0$ along the direction $\hat{\boldsymbol{b}}_0 \equiv \boldsymbol{B}_0 / \left\vert \boldsymbol{B}_0 \right\vert$, and we have neglected density fluctuations.
Since we are interested in the Alfvén-wave dynamics, we assume transverse and non-compressive fluctuations, i.e.,
\begin{subequations}
\begin{align}
    \delta \boldsymbol{u} \cdot \boldsymbol{B}_0 &= \delta \boldsymbol{B} \cdot \boldsymbol{B}_0 = 0, \\
    \boldsymbol{\nabla} \cdot \delta \boldsymbol{u} &= \boldsymbol{\nabla} \cdot \delta \boldsymbol{B} = 0,
\end{align}
\end{subequations}
and we take $\rho$, $U$ and $B_0$ to be steady-state solutions of (\ref{eq:continuity}) through (\ref{eq:induction}). The Alfvén velocity and \cite{Elsasser_1950_PhysRev} variables are respectively given by 
\begin{equation}
    \boldsymbol{v}_{\rm A} (\boldsymbol{r}) = \frac{\boldsymbol{B}_0 (\boldsymbol{r}) }{\sqrt{4 \pi \rho (\boldsymbol{r})}}, \quad \text{and} \quad \boldsymbol{z}^\pm (\boldsymbol{r}, t)= \delta \boldsymbol{u} (\boldsymbol{r}, t) \mp \delta \boldsymbol{b} (\boldsymbol{r}, t),
\end{equation}
where $\delta \boldsymbol{b} (\boldsymbol{r}, t) \equiv \delta \boldsymbol{B} (\boldsymbol{r}, t)/ \sqrt{4 \pi \rho (\boldsymbol{r})}$. 
We take $\boldsymbol{B}_0$ to be directed away from the Sun, so that $\boldsymbol{z}^+$ ($\boldsymbol{z}^-$) corresponds to outward-propagating (inward-propagating) fluctuations. 
Given these assumptions, the fluctuations are well described by the equations of IRMHD \citep{Velli_1989_PRL, Zhou_1989_GRL, Cranmer_2005_ApJSS, Verdini_2007_ApJ, Perez_2013_ApJ}:
\begin{equation} \label{eq:Elsasser}
    \frac{\partial \boldsymbol{z}^\pm}{\partial t} + \left( U \pm v_{\rm A} \right) \left( \hat{\boldsymbol{b}}_0 \cdot \boldsymbol{\nabla} \boldsymbol{z}^\pm \right) + \left( U \mp v_{\rm A} \right) \left( \frac{\boldsymbol{z}^\pm}{4 H_\rho} - \frac{\boldsymbol{z}^\mp}{2 H_{\rm A}} \right)
    = - \boldsymbol{z}^\mp \cdot \boldsymbol{\nabla} \boldsymbol{z}^\pm - \boldsymbol{\nabla} p_\star,
\end{equation}
where $p_\star = p_\mathrm{tot}/ \rho$, and the $-\boldsymbol{\nabla} p_\star$ term on the right-hand side of (\ref{eq:Elsasser}) enforces $ \boldsymbol{\nabla} \cdot \boldsymbol{z}^\pm = 0$.
The parameters $H_\rho(r)$, and $H_{\rm A}(r)$, are, respectively, the scale heights of the background mass density and the Alfvén speed. We also define the scale height of the background magnetic field $H_{\rm B}(r)$, giving
\begin{equation} \label{eq:height-scales}
    \frac{1}{H_{\rm B}} \equiv - \frac{\hat{\boldsymbol{b}}_0 \cdot \boldsymbol{\nabla} B_0}{B_0} > 0,
    \quad
    \frac{1}{H_\rho} \equiv - \frac{\hat{\boldsymbol{b}}_0 \cdot \boldsymbol{\nabla} \rho}{\rho} > 0,
    \quad
    \frac{1}{H_{\rm A}} \equiv \frac{\hat{\boldsymbol{b}}_0 \cdot \boldsymbol{\nabla} v_{\rm A}}{v_{\rm A}} = \frac{1}{2 H_\rho} - \frac{1}{H_{\rm B}}.
\end{equation}

Eqs. (\ref{eq:Elsasser}) are nicely expressed in terms of the wave-action variables introduced by \cite{Heinemann_1980_JGR}:  
\begin{equation}
    \label{eq:fpm}
    \boldsymbol{f}^\pm (\boldsymbol{r}, t) \equiv \frac{1 \pm \eta^{1/2}}{\eta^{1/4}} \boldsymbol{z}^\pm, \quad \text{where} \quad \eta (\boldsymbol{r}) \equiv \frac{\rho (\boldsymbol{r})}{\rho_{\rm A}},
\end{equation}
where $\rho_{\rm A}$ is the density at the Alfvén surface, which is defined as the surface on which $U = v_{\rm A}$
\footnote{The Alfvén surface is an idealization of what is really happening, and that there is more of a fuzzy transition region than a smooth and clean transition between sub- and super-Alfvénic solar wind \citep{Chhiber_2022_MNRAS}.}.
Upon substituting~(\ref{eq:height-scales}) and~(\ref{eq:fpm}) into~(\ref{eq:Elsasser}), we find that \citep{Heinemann_1980_JGR, Chandran_2009_ApJ, Chandran_2019_JPP}
\begin{equation} \label{eq:Heinemann-Olbert}
    \frac{\partial \boldsymbol{f}^\pm}{\partial t} + \left( U \pm v_{\rm A} \right) \left( \hat{\boldsymbol{b}}_0 \cdot \boldsymbol{\nabla} \boldsymbol{f}^\pm - \frac{\boldsymbol{f}^\mp}{2 H_{\rm A}} \right)
    = - \boldsymbol{z}^\mp \cdot \boldsymbol{\nabla} \boldsymbol{f}^\pm - \frac{1 \pm \eta^{1/2}}{\eta^{1/4}} \boldsymbol{\nabla} p_\star.
\end{equation}
Thanks to the conservation of magnetic flux and mass, $\rho U / B_0 = $  constant, and $\eta = \mathcal{M}_{\rm A}^{-2}$, where $\mathcal{M}_{\rm A} \equiv U/v_{\rm A}$ is the Alfvénic Mach number.
In the WKB (Wentzel–Kramers–Brillouin) limit, and in the absence of nonlinear interactions between counter-propagating waves, the average of $\left(\boldsymbol{f}^+\right)^2$ over a wave period is independent of $r$ \citep{Heinemann_1980_JGR}, and hence, because of Eqs. (\ref{eq:fpm}), the average of $\left(\boldsymbol{z}^+\right)^2$ over a wave period peaks at the Alfvén critical point, where $\mathcal{M}_{\rm A}=1$.

%-------------------------------
\subsection{Clebsch coordinates}
\label{sec:Clebsch}

\begin{figure}
    \centering
    
    \tikzset{every picture/.style={line width=0.75pt}} %set default line width to 0.75pt        
    
    \begin{tikzpicture}[x=0.75pt,y=0.75pt,yscale=-1,xscale=1]
    
    %Curve Lines [id:da32875139961834554] 
    \draw [line width=1.5]    (56.53,48.76) .. controls (65.57,34.8) and (117.2,14.62) .. (162.24,6.41) ;
    \draw [shift={(165.69,5.8)}, rotate = 170.36] [fill={rgb, 255:red, 0; green, 0; blue, 0 }  ][line width=0.08]  [draw opacity=0] (4.64,-2.23) -- (0,0) -- (4.64,2.23) -- cycle    ;
    %Curve Lines [id:da09244233088774956] 
    \draw [line width=1.5]    (56.53,48.76) .. controls (69.7,58.22) and (124.55,59.84) .. (170.19,52.08) ;
    \draw [shift={(173.68,51.47)}, rotate = 169.65] [fill={rgb, 255:red, 0; green, 0; blue, 0 }  ][line width=0.08]  [draw opacity=0] (4.64,-2.23) -- (0,0) -- (4.64,2.23) -- cycle    ;
    %Curve Lines [id:da615937098012052] 
    \draw [line width=0.75]    (56.53,48.76) .. controls (76.37,33.41) and (121.35,25.35) .. (166.26,17.71) ;
    \draw [shift={(169.01,17.24)}, rotate = 170.36] [fill={rgb, 255:red, 0; green, 0; blue, 0 }  ][line width=0.08]  [draw opacity=0] (3.57,-1.72) -- (0,0) -- (3.57,1.72) -- cycle    ;
    %Curve Lines [id:da8793423674516185] 
    \draw [line width=0.75]    (56.53,48.76) .. controls (80.89,55.74) and (125.26,48.31) .. (170.59,40.05) ;
    \draw [shift={(173.36,39.54)}, rotate = 169.65] [fill={rgb, 255:red, 0; green, 0; blue, 0 }  ][line width=0.08]  [draw opacity=0] (3.57,-1.72) -- (0,0) -- (3.57,1.72) -- cycle    ;
    %Straight Lines [id:da3709150712364624] 
    \draw [color={rgb, 255:red, 0; green, 64; blue, 221 }  ,draw opacity=1 ][line width=0.75]    (56.53,48.76) -- (174.96,27.84) ;
    \draw [shift={(177.92,27.32)}, rotate = 169.98] [fill={rgb, 255:red, 0; green, 64; blue, 221 }  ,fill opacity=1 ][line width=0.08]  [draw opacity=0] (5.36,-2.57) -- (0,0) -- (5.36,2.57) -- cycle    ;
    %Shape: Circle [id:dp36265872742591787] 
    \draw  [color={rgb, 255:red, 0; green, 0; blue, 0 }  ,draw opacity=1 ][fill={rgb, 255:red, 0; green, 0; blue, 0 }  ,fill opacity=1 ] (7.5,54) .. controls (7.5,40.19) and (18.69,29) .. (32.5,29) .. controls (46.31,29) and (57.5,40.19) .. (57.5,54) .. controls (57.5,67.81) and (46.31,79) .. (32.5,79) .. controls (18.69,79) and (7.5,67.81) .. (7.5,54) -- cycle ;
    %Curve Lines [id:da9020447971897816] 
    \draw [line width=0.75]    (56.53,48.76) .. controls (80.89,55.74) and (125.5,55.02) .. (170.84,47.03) ;
    \draw [shift={(173.62,46.53)}, rotate = 169.65] [fill={rgb, 255:red, 0; green, 0; blue, 0 }  ][line width=0.08]  [draw opacity=0] (3.57,-1.72) -- (0,0) -- (3.57,1.72) -- cycle    ;
    %Curve Lines [id:da3467697500919107] 
    \draw [line width=0.75]    (56.53,48.76) .. controls (76.37,33.41) and (119.47,19.11) .. (164.31,11.21) ;
    \draw [shift={(167.06,10.74)}, rotate = 170.36] [fill={rgb, 255:red, 0; green, 0; blue, 0 }  ][line width=0.08]  [draw opacity=0] (3.57,-1.72) -- (0,0) -- (3.57,1.72) -- cycle    ;
    %Curve Lines [id:da6749122159742587] 
    \draw [line width=0.75]    (56.53,48.76) .. controls (76.71,38.45) and (123.23,31.8) .. (168.21,24.21) ;
    \draw [shift={(170.96,23.74)}, rotate = 170.36] [fill={rgb, 255:red, 0; green, 0; blue, 0 }  ][line width=0.08]  [draw opacity=0] (3.57,-1.72) -- (0,0) -- (3.57,1.72) -- cycle    ;
    %Curve Lines [id:da5153214178849541] 
    \draw [line width=0.75]    (56.53,48.76) .. controls (81.17,51.52) and (124.26,42.36) .. (169.54,34.03) ;
    \draw [shift={(172.32,33.52)}, rotate = 169.65] [fill={rgb, 255:red, 0; green, 0; blue, 0 }  ][line width=0.08]  [draw opacity=0] (3.57,-1.72) -- (0,0) -- (3.57,1.72) -- cycle    ;
    %Straight Lines [id:da14383251717608325] 
    \draw [color={rgb, 255:red, 0; green, 122; blue, 255 }  ,draw opacity=1 ]   (94.32,104.05) -- (299.21,81.16) ;
    \draw [shift={(198.16,92.45)}, rotate = 173.63] [fill={rgb, 255:red, 0; green, 122; blue, 255 }  ,fill opacity=1 ][line width=0.08]  [draw opacity=0] (3.57,-1.72) -- (0,0) -- (3.57,1.72) -- cycle    ;
    %Straight Lines [id:da12163639673168536] 
    \draw [color={rgb, 255:red, 0; green, 122; blue, 255 }  ,draw opacity=1 ]   (94.3,103.94) -- (295.5,54.9) ;
    \draw [shift={(196.26,79.09)}, rotate = 166.3] [fill={rgb, 255:red, 0; green, 122; blue, 255 }  ,fill opacity=1 ][line width=0.08]  [draw opacity=0] (3.57,-1.72) -- (0,0) -- (3.57,1.72) -- cycle    ;
    %Straight Lines [id:da7010219732903078] 
    \draw [color={rgb, 255:red, 0; green, 122; blue, 255 }  ,draw opacity=1 ]   (94.32,104.05) -- (293.04,63.46) ;
    \draw [shift={(195.05,83.47)}, rotate = 168.46] [fill={rgb, 255:red, 0; green, 122; blue, 255 }  ,fill opacity=1 ][line width=0.08]  [draw opacity=0] (3.57,-1.72) -- (0,0) -- (3.57,1.72) -- cycle    ;
    %Straight Lines [id:da07434233993616912] 
    \draw [color={rgb, 255:red, 0; green, 122; blue, 255 }  ,draw opacity=1 ]   (94.3,103.94) -- (295.26,73.73) ;
    \draw [shift={(196.17,88.62)}, rotate = 171.45] [fill={rgb, 255:red, 0; green, 122; blue, 255 }  ,fill opacity=1 ][line width=0.08]  [draw opacity=0] (3.57,-1.72) -- (0,0) -- (3.57,1.72) -- cycle    ;
    %Straight Lines [id:da2979711202091375] 
    \draw [color={rgb, 255:red, 0; green, 64; blue, 221 }  ,draw opacity=1 ]   (94.41,104.54) -- (310.8,66.03) ;
    \draw [shift={(313.75,65.5)}, rotate = 169.91] [fill={rgb, 255:red, 0; green, 64; blue, 221 }  ,fill opacity=1 ][line width=0.08]  [draw opacity=0] (3.57,-1.72) -- (0,0) -- (3.57,1.72) -- cycle    ;
    %Straight Lines [id:da1911296172224708] 
    \draw [color={rgb, 255:red, 0; green, 0; blue, 0 }  ,draw opacity=1 ]   (94.41,104.54) -- (294.48,94.09) ;
    %Straight Lines [id:da8598721774953908] 
    \draw [color={rgb, 255:red, 0; green, 0; blue, 0 }  ,draw opacity=1 ]   (94.41,104.54) -- (292.98,40.73) ;
    %Straight Lines [id:da30845649644149753] 
    \draw [color={rgb, 255:red, 0; green, 0; blue, 0 }  ,draw opacity=1 ]   (94.41,104.54) -- (287.6,52.77) ;
    %Straight Lines [id:da828514012883971] 
    \draw [color={rgb, 255:red, 0; green, 0; blue, 0 }  ,draw opacity=1 ] [dash pattern={on 0.84pt off 2.51pt}]  (97.98,103.91) -- (302.79,85.68) ;
    %Shape: Arc [id:dp48204100314415244] 
    \draw  [draw opacity=0] (287.6,52.77) .. controls (291.2,66.26) and (293.41,80.1) .. (294.15,94.07) -- (94.41,104.54) -- cycle ; \draw  [color={rgb, 255:red, 0; green, 0; blue, 0 }  ,draw opacity=1 ] (287.6,52.77) .. controls (291.2,66.26) and (293.41,80.1) .. (294.15,94.07) ;  
    %Shape: Arc [id:dp0720555515804242] 
    \draw  [draw opacity=0] (293.72,40.38) .. controls (298.39,55.09) and (301.43,70.29) .. (302.79,85.68) -- (93.58,103.98) -- cycle ; \draw  [color={rgb, 255:red, 0; green, 0; blue, 0 }  ,draw opacity=1 ] (293.72,40.38) .. controls (298.39,55.09) and (301.43,70.29) .. (302.79,85.68) ;  
    %Shape: Arc [id:dp8967263161227812] 
    \draw  [draw opacity=0] (292.98,40.73) .. controls (292.98,40.73) and (292.98,40.73) .. (292.98,40.73) .. controls (293.87,44.92) and (291.5,49.43) .. (286.85,53.47) -- (258.75,48.01) -- cycle ; \draw  [color={rgb, 255:red, 0; green, 0; blue, 0 }  ,draw opacity=1 ] (292.98,40.73) .. controls (292.98,40.73) and (292.98,40.73) .. (292.98,40.73) .. controls (293.87,44.92) and (291.5,49.43) .. (286.85,53.47) ;  
    %Shape: Arc [id:dp8585980144281324] 
    \draw  [draw opacity=0] (302.63,85.44) .. controls (303.17,87.99) and (299.9,91.1) .. (294.15,94.07) -- (268.4,92.72) -- cycle ; \draw  [color={rgb, 255:red, 0; green, 0; blue, 0 }  ,draw opacity=1 ] (302.63,85.44) .. controls (303.17,87.99) and (299.9,91.1) .. (294.15,94.07) ;  
    
    %Shape: Cube [id:dp3378327846978868] 
    \draw   (348.25,28.5) -- (369.25,7.5) -- (461.25,7.5) -- (461.25,56.5) -- (440.25,77.5) -- (348.25,77.5) -- cycle ; \draw   (461.25,7.5) -- (440.25,28.5) -- (348.25,28.5) ; \draw   (440.25,28.5) -- (440.25,77.5) ;
    %Straight Lines [id:da8006815028198249] 
    \draw [color={rgb, 255:red, 0; green, 64; blue, 221 }  ,draw opacity=1 ]   (466.75,104.5) -- (484.63,86.62) ;
    \draw [shift={(486.75,84.5)}, rotate = 135] [fill={rgb, 255:red, 0; green, 64; blue, 221 }  ,fill opacity=1 ][line width=0.08]  [draw opacity=0] (3.57,-1.72) -- (0,0) -- (3.57,1.72) -- cycle    ;
    %Straight Lines [id:da32384258865648485] 
    \draw [color={rgb, 255:red, 0; green, 64; blue, 221 }  ,draw opacity=1 ]   (466.75,104.5) -- (483.75,104.5) ;
    \draw [shift={(486.75,104.5)}, rotate = 180] [fill={rgb, 255:red, 0; green, 64; blue, 221 }  ,fill opacity=1 ][line width=0.08]  [draw opacity=0] (3.57,-1.72) -- (0,0) -- (3.57,1.72) -- cycle    ;
    %Straight Lines [id:da022206855794547842] 
    \draw [color={rgb, 255:red, 0; green, 64; blue, 221 }  ,draw opacity=1 ]   (466.75,104.5) -- (466.75,87.5) ;
    \draw [shift={(466.75,84.5)}, rotate = 90] [fill={rgb, 255:red, 0; green, 64; blue, 221 }  ,fill opacity=1 ][line width=0.08]  [draw opacity=0] (3.57,-1.72) -- (0,0) -- (3.57,1.72) -- cycle    ;
    
    %Straight Lines [id:da8134135410700511] 
    \draw [color={rgb, 255:red, 0; green, 122; blue, 255 }  ,draw opacity=1 ]   (352.5,70.5) -- (442.5,70.5) ;
    \draw [shift={(398.9,70.5)}, rotate = 180] [fill={rgb, 255:red, 0; green, 122; blue, 255 }  ,fill opacity=1 ][line width=0.08]  [draw opacity=0] (3.57,-1.72) -- (0,0) -- (3.57,1.72) -- cycle    ;
    %Straight Lines [id:da6025034252686023] 
    \draw [color={rgb, 255:red, 0; green, 122; blue, 255 }  ,draw opacity=1 ]   (368.5,50) -- (456.5,50) ;
    \draw [shift={(413.9,50)}, rotate = 180] [fill={rgb, 255:red, 0; green, 122; blue, 255 }  ,fill opacity=1 ][line width=0.08]  [draw opacity=0] (3.57,-1.72) -- (0,0) -- (3.57,1.72) -- cycle    ;
    %Straight Lines [id:da8401127646290403] 
    \draw [color={rgb, 255:red, 0; green, 122; blue, 255 }  ,draw opacity=1 ]   (359,34.5) -- (449,34.5) ;
    \draw [shift={(405.4,34.5)}, rotate = 180] [fill={rgb, 255:red, 0; green, 122; blue, 255 }  ,fill opacity=1 ][line width=0.08]  [draw opacity=0] (3.57,-1.72) -- (0,0) -- (3.57,1.72) -- cycle    ;
    %Straight Lines [id:da8907092890499093] 
    \draw [color={rgb, 255:red, 0; green, 122; blue, 255 }  ,draw opacity=1 ]   (349.5,62) -- (442.5,62) ;
    \draw [shift={(397.4,62)}, rotate = 180] [fill={rgb, 255:red, 0; green, 122; blue, 255 }  ,fill opacity=1 ][line width=0.08]  [draw opacity=0] (3.57,-1.72) -- (0,0) -- (3.57,1.72) -- cycle    ;
    %Straight Lines [id:da8501961204133908] 
    \draw [color={rgb, 255:red, 0; green, 122; blue, 255 }  ,draw opacity=1 ]   (350,43) -- (443,43) ;
    \draw [shift={(397.9,43)}, rotate = 180] [fill={rgb, 255:red, 0; green, 122; blue, 255 }  ,fill opacity=1 ][line width=0.08]  [draw opacity=0] (3.57,-1.72) -- (0,0) -- (3.57,1.72) -- cycle    ;
    %Straight Lines [id:da3698344012884158] 
    \draw [color={rgb, 255:red, 0; green, 122; blue, 255 }  ,draw opacity=1 ]   (353.75,55.5) -- (446.5,55.5) ;
    \draw [shift={(401.53,55.5)}, rotate = 180] [fill={rgb, 255:red, 0; green, 122; blue, 255 }  ,fill opacity=1 ][line width=0.08]  [draw opacity=0] (3.57,-1.72) -- (0,0) -- (3.57,1.72) -- cycle    ;
    %Curve Lines [id:da426034444304614] 
    \draw [color={rgb, 255:red, 0; green, 122; blue, 255 }  ,draw opacity=0.7 ][line width=0.75]    (57.03,164.76) .. controls (66.12,150.73) and (118.23,130.42) .. (163.43,122.28) ;
    \draw [shift={(166.19,121.8)}, rotate = 170.36] [fill={rgb, 255:red, 0; green, 122; blue, 255 }  ,fill opacity=0.7 ][line width=0.08]  [draw opacity=0] (3.57,-1.72) -- (0,0) -- (3.57,1.72) -- cycle    ;
    %Curve Lines [id:da48260703213995715] 
    \draw [color={rgb, 255:red, 0; green, 122; blue, 255 }  ,draw opacity=0.7 ][line width=0.75]    (57.03,164.76) .. controls (70.27,174.27) and (125.62,175.86) .. (171.39,167.96) ;
    \draw [shift={(174.18,167.47)}, rotate = 169.65] [fill={rgb, 255:red, 0; green, 122; blue, 255 }  ,fill opacity=0.7 ][line width=0.08]  [draw opacity=0] (3.57,-1.72) -- (0,0) -- (3.57,1.72) -- cycle    ;
    %Curve Lines [id:da4074726695191524] 
    \draw [color={rgb, 255:red, 0; green, 122; blue, 255 }  ,draw opacity=0.7 ][line width=0.75]    (57.03,164.76) .. controls (76.87,149.41) and (121.85,141.35) .. (166.76,133.71) ;
    \draw [shift={(169.51,133.24)}, rotate = 170.36] [fill={rgb, 255:red, 0; green, 122; blue, 255 }  ,fill opacity=0.7 ][line width=0.08]  [draw opacity=0] (3.57,-1.72) -- (0,0) -- (3.57,1.72) -- cycle    ;
    %Curve Lines [id:da6538604147489055] 
    \draw [color={rgb, 255:red, 0; green, 122; blue, 255 }  ,draw opacity=0.7 ][line width=0.75]    (57.03,164.76) .. controls (81.39,171.74) and (125.76,164.31) .. (171.09,156.05) ;
    \draw [shift={(173.86,155.54)}, rotate = 169.65] [fill={rgb, 255:red, 0; green, 122; blue, 255 }  ,fill opacity=0.7 ][line width=0.08]  [draw opacity=0] (3.57,-1.72) -- (0,0) -- (3.57,1.72) -- cycle    ;
    %Straight Lines [id:da04098841794818253] 
    \draw [color={rgb, 255:red, 0; green, 64; blue, 221 }  ,draw opacity=0.7 ][line width=0.75]    (57.03,164.76) -- (175.46,143.84) ;
    \draw [shift={(178.42,143.32)}, rotate = 169.98] [fill={rgb, 255:red, 0; green, 64; blue, 221 }  ,fill opacity=0.7 ][line width=0.08]  [draw opacity=0] (5.36,-2.57) -- (0,0) -- (5.36,2.57) -- cycle    ;
    %Straight Lines [id:da30272913599311013] 
    \draw [line width=0.75]  [dash pattern={on 3.75pt off 1.5pt}]  (106.61,141.53) -- (106.61,179.14) ;
    %Straight Lines [id:da6063842619581012] 
    \draw [line width=0.75]  [dash pattern={on 3.75pt off 1.5pt}]  (122.73,125.41) -- (122.73,163.02) ;
    %Straight Lines [id:da3543594539868431] 
    \draw [line width=0.75]  [dash pattern={on 3.75pt off 1.5pt}]  (122.73,163.02) -- (106.61,179.14) ;
    %Straight Lines [id:da3406706558214647] 
    \draw [line width=0.75]  [dash pattern={on 3.75pt off 1.5pt}]  (122.73,125.41) -- (106.61,141.53) ;
    
    %Curve Lines [id:da6693898969378649] 
    \draw [color={rgb, 255:red, 0; green, 122; blue, 255 }  ,draw opacity=0.7 ][line width=0.75]    (257.53,158.62) .. controls (266.62,144.59) and (318.73,124.28) .. (363.93,116.14) ;
    \draw [shift={(366.69,115.66)}, rotate = 170.36] [fill={rgb, 255:red, 0; green, 122; blue, 255 }  ,fill opacity=0.7 ][line width=0.08]  [draw opacity=0] (3.57,-1.72) -- (0,0) -- (3.57,1.72) -- cycle    ;
    %Curve Lines [id:da3449888239282164] 
    \draw [color={rgb, 255:red, 0; green, 122; blue, 255 }  ,draw opacity=0.7 ][line width=0.75]    (257.53,158.62) .. controls (270.77,168.13) and (326.12,169.72) .. (371.89,161.82) ;
    \draw [shift={(374.68,161.33)}, rotate = 169.65] [fill={rgb, 255:red, 0; green, 122; blue, 255 }  ,fill opacity=0.7 ][line width=0.08]  [draw opacity=0] (3.57,-1.72) -- (0,0) -- (3.57,1.72) -- cycle    ;
    %Curve Lines [id:da11296092714367956] 
    \draw [color={rgb, 255:red, 0; green, 122; blue, 255 }  ,draw opacity=0.7 ][line width=0.75]    (257.53,158.62) .. controls (277.37,143.27) and (322.35,135.21) .. (367.26,127.57) ;
    \draw [shift={(370.01,127.1)}, rotate = 170.36] [fill={rgb, 255:red, 0; green, 122; blue, 255 }  ,fill opacity=0.7 ][line width=0.08]  [draw opacity=0] (3.57,-1.72) -- (0,0) -- (3.57,1.72) -- cycle    ;
    %Curve Lines [id:da17575185385795544] 
    \draw [color={rgb, 255:red, 0; green, 122; blue, 255 }  ,draw opacity=0.7 ][line width=0.75]    (257.53,158.62) .. controls (281.89,165.6) and (326.26,158.17) .. (371.59,149.91) ;
    \draw [shift={(374.36,149.41)}, rotate = 169.65] [fill={rgb, 255:red, 0; green, 122; blue, 255 }  ,fill opacity=0.7 ][line width=0.08]  [draw opacity=0] (3.57,-1.72) -- (0,0) -- (3.57,1.72) -- cycle    ;
    %Straight Lines [id:da724538867097025] 
    \draw [color={rgb, 255:red, 0; green, 64; blue, 221 }  ,draw opacity=0.7 ][line width=0.75]    (257.53,158.62) -- (375.96,137.7) ;
    \draw [shift={(378.92,137.18)}, rotate = 169.98] [fill={rgb, 255:red, 0; green, 64; blue, 221 }  ,fill opacity=0.7 ][line width=0.08]  [draw opacity=0] (5.36,-2.57) -- (0,0) -- (5.36,2.57) -- cycle    ;
    %Straight Lines [id:da12829296804027335] 
    \draw [line width=0.75]  [dash pattern={on 3.75pt off 1.5pt}]  (326.11,129.91) -- (326.11,170.5) ;
    %Straight Lines [id:da7117751695799401] 
    \draw [line width=0.75]  [dash pattern={on 3.75pt off 1.5pt}]  (343.5,112.52) -- (343.5,153.11) ;
    %Straight Lines [id:da9022374675121843] 
    \draw [line width=0.75]  [dash pattern={on 3.75pt off 1.5pt}]  (343.5,153.11) -- (326.11,170.5) ;
    %Straight Lines [id:da6844656467272223] 
    \draw [line width=0.75]  [dash pattern={on 3.75pt off 1.5pt}]  (343.5,112.52) -- (326.11,129.91) ;
    
    %Straight Lines [id:da29390078663881813] 
    \draw [color={rgb, 255:red, 0; green, 0; blue, 0 }  ,draw opacity=0.3 ][line width=0.75]  [dash pattern={on 3.75pt off 1.5pt}]  (276.11,144.67) -- (276.11,169) ;
    %Straight Lines [id:da2835273920304453] 
    \draw [color={rgb, 255:red, 0; green, 0; blue, 0 }  ,draw opacity=0.3 ][line width=0.75]  [dash pattern={on 3.75pt off 1.5pt}]  (286.53,134.25) -- (286.53,158.57) ;
    %Straight Lines [id:da21420946227996107] 
    \draw [color={rgb, 255:red, 0; green, 0; blue, 0 }  ,draw opacity=0.3 ][line width=0.75]  [dash pattern={on 3.75pt off 1.5pt}]  (286.53,158.57) -- (276.11,169) ;
    %Straight Lines [id:da4080710514468331] 
    \draw [color={rgb, 255:red, 0; green, 0; blue, 0 }  ,draw opacity=0.3 ][line width=0.75]  [dash pattern={on 3.75pt off 1.5pt}]  (286.53,134.25) -- (276.11,144.67) ;
    
    %Straight Lines [id:da14375984545897813] 
    \draw [color={rgb, 255:red, 0; green, 0; blue, 0 }  ,draw opacity=0.3 ][line width=0.75]  [dash pattern={on 3.75pt off 1.5pt}]  (291.61,140.42) -- (291.61,170) ;
    %Straight Lines [id:da4428191764703435] 
    \draw [color={rgb, 255:red, 0; green, 0; blue, 0 }  ,draw opacity=0.3 ][line width=0.75]  [dash pattern={on 3.75pt off 1.5pt}]  (304.28,127.75) -- (304.28,157.32) ;
    %Straight Lines [id:da7654109701402019] 
    \draw [color={rgb, 255:red, 0; green, 0; blue, 0 }  ,draw opacity=0.3 ][line width=0.75]  [dash pattern={on 3.75pt off 1.5pt}]  (304.28,157.32) -- (291.61,170) ;
    %Straight Lines [id:da22577256494153364] 
    \draw [color={rgb, 255:red, 0; green, 0; blue, 0 }  ,draw opacity=0.3 ][line width=0.75]  [dash pattern={on 3.75pt off 1.5pt}]  (304.28,127.75) -- (291.61,140.42) ;
    
    %Straight Lines [id:da4180383185684775] 
    \draw [color={rgb, 255:red, 0; green, 0; blue, 0 }  ,draw opacity=0.3 ][line width=0.75]  [dash pattern={on 3.75pt off 1.5pt}]  (308.38,135.45) -- (308.38,169.75) ;
    %Straight Lines [id:da9262730015190834] 
    \draw [color={rgb, 255:red, 0; green, 0; blue, 0 }  ,draw opacity=0.3 ][line width=0.75]  [dash pattern={on 3.75pt off 1.5pt}]  (323.08,120.75) -- (323.08,155.05) ;
    %Straight Lines [id:da012036862405119475] 
    \draw [color={rgb, 255:red, 0; green, 0; blue, 0 }  ,draw opacity=0.3 ][line width=0.75]  [dash pattern={on 3.75pt off 1.5pt}]  (323.08,155.05) -- (308.38,169.75) ;
    %Straight Lines [id:da650935206837209] 
    \draw [color={rgb, 255:red, 0; green, 0; blue, 0 }  ,draw opacity=0.3 ][line width=0.75]  [dash pattern={on 3.75pt off 1.5pt}]  (323.08,120.75) -- (308.38,135.45) ;
    
    %Shape: Circle [id:dp5454483770791205] 
    \draw  [color={rgb, 255:red, 0; green, 0; blue, 0 }  ,draw opacity=1 ][fill={rgb, 255:red, 0; green, 0; blue, 0 }  ,fill opacity=1 ] (7.03,164.76) .. controls (7.03,150.95) and (18.22,139.76) .. (32.03,139.76) .. controls (45.84,139.76) and (57.03,150.95) .. (57.03,164.76) .. controls (57.03,178.56) and (45.84,189.76) .. (32.03,189.76) .. controls (18.22,189.76) and (7.03,178.56) .. (7.03,164.76) -- cycle ;
    %Shape: Rectangle [id:dp5571513982632441] 
    \draw  [draw opacity=0][fill={rgb, 255:red, 255; green, 255; blue, 255 }  ,fill opacity=1 ] (0,137.25) -- (45.5,137.25) -- (45.5,193) -- (0,193) -- cycle ;
    %Shape: Circle [id:dp07622279967366197] 
    \draw  [color={rgb, 255:red, 0; green, 0; blue, 0 }  ,draw opacity=1 ][fill={rgb, 255:red, 0; green, 0; blue, 0 }  ,fill opacity=1 ] (207.75,159.63) .. controls (207.75,145.82) and (218.94,134.63) .. (232.75,134.63) .. controls (246.56,134.63) and (257.75,145.82) .. (257.75,159.63) .. controls (257.75,173.43) and (246.56,184.63) .. (232.75,184.63) .. controls (218.94,184.63) and (207.75,173.43) .. (207.75,159.63) -- cycle ;
    %Shape: Rectangle [id:dp20421820028272197] 
    \draw  [draw opacity=0][fill={rgb, 255:red, 255; green, 255; blue, 255 }  ,fill opacity=1 ] (199,129.25) -- (244.5,129.25) -- (244.5,185) -- (199,185) -- cycle ;
    
    % Text Node
    \draw (32.5,54) node  [color={rgb, 255:red, 255; green, 255; blue, 255 }  ,opacity=1 ] [align=left] {Sun};
    % Text Node
    \draw (315.75,68.9) node [anchor=north west][inner sep=0.75pt]    {$\hat{\boldsymbol{e}}_{r}$};
    % Text Node
    \draw (486.75,107.9) node [anchor=north] [inner sep=0.75pt]    {$\hat{\boldsymbol{e}}_{\tilde{z}}$};
    % Text Node
    \draw (488.75,84.5) node [anchor=west] [inner sep=0.75pt]    {$\hat{\boldsymbol{e}}_{\tilde{x}}$};
    % Text Node
    \draw (464.75,84.5) node [anchor=east] [inner sep=0.75pt]    {$\hat{\boldsymbol{e}}_{\tilde{y}}$};
    % Text Node
    \draw (2.67,4) node [anchor=north west][inner sep=0.75pt]   [align=left] {a)};
    % Text Node
    \draw (92,77) node [anchor=north west][inner sep=0.75pt]   [align=left] {b)};
    % Text Node
    \draw (328.67,7.67) node [anchor=north west][inner sep=0.75pt]   [align=left] {c)};
    % Text Node
    \draw (106.61,182.54) node [anchor=north] [inner sep=0.75pt]    {$\boldsymbol{U}_{\rm fr} =\boldsymbol{0}$};
    % Text Node
    \draw (51,126) node [anchor=north west][inner sep=0.75pt]   [align=left] {d)};
    % Text Node
    \draw (248.5,116.83) node [anchor=north west][inner sep=0.75pt]   [align=left] {e)};
    % Text Node
    \draw (322.61,178.9) node [anchor=north east] [inner sep=0.75pt]    {$\boldsymbol{U}_{\rm fr} \neq \boldsymbol{0}$};

    \end{tikzpicture}

    \caption{a) A cartoon of a magnetic flux tube expanding from the Sun. The dark arrows represent magnetic field lines, and the blue arrow shows the central field line of the flux tube.
    b) A model of a narrow magnetic flux tube.
    c) The same narrow magnetic flux tube after applying Clebsch coordinates.
    d) Cartoon of the Eulerian box simulation: the box remains fixed while the plasma flows through it. 
    e) Cartoon of the moving box simulation: the box moves along with the plasma at a specified velocity, expanding with the field lines.}
    \label{fig:flux-tube}
\end{figure}
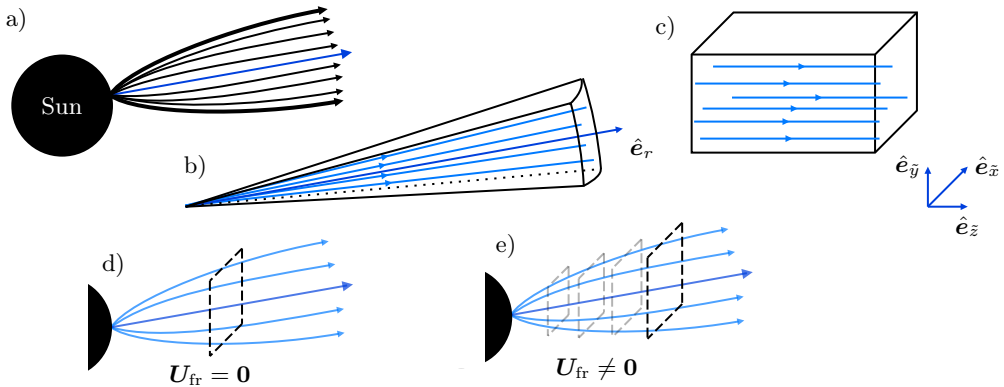

To facilitate the design of future numerical simulations, we introduce \cite{Clebsch_1871} coordinates $\tilde{x}$ and $\tilde{y}$ that  are constant along the background magnetic-field lines.
In particular, we define
\begin{equation} \label{eq:Clebsch}
  (\tilde{x}, \tilde{y}, \tilde{z})=(\alpha x, \alpha y, z),
\end{equation} where $\alpha^2 \equiv \bar{B}_0/B_\text{ref}$, and $B_\text{ref}$ is a value of reference.
The function $\alpha(r)$ is the factor by which the separation between two field lines increases between radius $r_{\rm ref}$ and some larger radius $r$.
These coordinates satisfy the relation 
\begin{equation} \label{eq:Clebsch2}
\frac{\boldsymbol{B}_0}{B_\text{ref}} = \boldsymbol{\nabla} \tilde{x} \times \boldsymbol{\nabla} \tilde{y} + O(\theta^2),
\end{equation}
from which it follows that $\boldsymbol{B}_0 \cdot \boldsymbol{\nabla} \tilde{x} = \boldsymbol{B}_0 \cdot \boldsymbol{\nabla} \tilde{y} = 0$.
 In the $(\tilde{x}, \tilde{y}, \tilde{z})$ coordinate system, Eqs.(\ref{eq:Elsasser}) and Eqs.(\ref{eq:Heinemann-Olbert}) transform into
\begin{equation} \label{eq:IRMHD-Clebsch-Elsasser}
    \frac{\partial \boldsymbol{z}^\pm}{\partial t} + \left( U \pm v_A \right) \frac{\partial \boldsymbol{z}^\pm}{\partial \tilde{z}} + \left( U \mp v_A \right) \left( \frac{\boldsymbol{z}^\pm}{4 H_\rho} - \frac{\boldsymbol{z}^\mp}{2 H_A} \right)
    = - \alpha \left( \boldsymbol{z}^\mp \cdot \tilde{\boldsymbol{\nabla}} \boldsymbol{z}^\pm +  \tilde{\boldsymbol{\nabla}}_\perp p_\star \right),
\end{equation}
and
\begin{equation} \label{eq:Clebsch-Heinemann-Olbert}
    \frac{\partial \boldsymbol{f}^\pm}{\partial t} + \left( U \pm v_A \right) \left( \frac{\partial \boldsymbol{f}^\pm}{\partial \tilde{z}} - \frac{\boldsymbol{f}^\mp}{2 H_A} \right)
    = - \alpha \left( \boldsymbol{z}^\mp \cdot \tilde{\boldsymbol{\nabla}} \boldsymbol{f}^\pm + \frac{1 \pm \eta^{1/2}}{\eta^{1/4}} \tilde{\boldsymbol{\nabla}}_\perp p_\star \right),
\end{equation}
respectively, where $\tilde{\boldsymbol{\nabla}}$ is the gradient operator in the $(\tilde{x}, \tilde{y}, \tilde{z})$ coordinate system.
In this coordinate system, the background magnetic-field lines are parallel to each other, as illustrated in Figure \ref{fig:flux-tube}.
The parameter $\alpha$ represents the separation distance between two field lines and adjusts the strength of the nonlinearities in the system.
For the sake of readability, we will henceforth employ the Clebsch coordinates exclusively, and omit the tilde notation.

%-----------------------------------------------------------------------
\subsection{Transformation to a frame moving radially away from the Sun}
\label{subsec:transformation}

In this section, we consider a reference frame that moves away from the Sun along the background magnetic-field lines at some arbitrary velocity $U_{\rm fr}(z)$. The position $z(t)$ of a point that
starts at $z_0$ at time $t_0$ and moves outward along $\boldsymbol{B}_0$ at speed $U_{\rm fr}(z)$ satisfies the equation
\begin{equation}
\label{eq:z0toz}
\frac{\partial}{\partial t} z \left(t \, | \, z_0, t_0 \right) = U_\mathrm{fr} \left[ z \left(t \, | \, z_0, t_0 \right) \right].
\end{equation}
In simpler words, this mapping relates the time elapsed since the initial moment $t_0$ to the distance $(z-z_0)$ traveled by the frame.
With the aid of (\ref{eq:z0toz}), we can transform (\ref{eq:Clebsch-Heinemann-Olbert}) from the $(x,y,z)$ coordinate system to the $(x,y,z_0)$ coordinate system, obtaining
\begin{equation} \label{eq:Heinemann-Olbert-Lagrangian}
    \frac{\partial \boldsymbol{f}^\pm}{\partial t} + \left( U - U_\mathrm{fr} \pm v_{\rm A} \right) \frac{U_{\mathrm{fr},0}}{U_\mathrm{fr}} \frac{\partial \boldsymbol{f}^\pm}{\partial z_0} - \frac{U \pm v_{\rm A}}{2 H_{\rm A}} \boldsymbol{f}^\mp = - \alpha \left( 
    \boldsymbol{z}^\mp \cdot \boldsymbol{\nabla} \boldsymbol{f}^\pm + \frac{1 \pm \eta^{1/2}}{\eta^{1/4}} \boldsymbol{\nabla} p_\star \right),
\end{equation}
where $U_{\mathrm{fr}} $ is shorthand for $U_{\mathrm{fr}}(z(z_0,t))$
and $U_{\mathrm{fr,0}} = U_{\mathrm{fr}}(z_0)$
The interested reader can find the details of the transformation calculations in Appendix \ref{app:Lagrangian}.

Eqs. (\ref{eq:Heinemann-Olbert-Lagrangian}) have two main differences compared with Eqs. (\ref{eq:Heinemann-Olbert}).
First, there is a correction to the advection term: $U \partial_z \to (U - U_\mathrm{fr}) \partial_z$.
This is intuitive because for a fixed observer ($U_\mathrm{fr} = 0$), we recover the usual advection term seen in the Eulerian frame.
As the frame speed $U_\mathrm{fr}$ gets closer to the plasma rest frame, the advection term decreases.
Second, we must account for the changes in the $z$-derivative due to the motion of the frame.
In general, as~$U_{\rm fr}$ depends on~$z$, the coordinate system will be stretched or compressed.
This deformation is captured by the transformation rule $\partial/\partial z \rightarrow (U_{\mathrm{fr,0}}/U_{\mathrm{fr}}) \partial/\partial z_0$.

Eqs. (\ref{eq:Heinemann-Olbert-Lagrangian}) represent the first main result of this paper.
The main advantage of this formulation is its flexibility in choosing the appropriate frame of reference.
For instance, in the plasma rest frame and the super-Alfvénic limit ($U_{\mathrm{fr}} = U_{\mathrm{fr},0} = U \simeq \text{constant}$ and $v_{\rm A} \ll U$), we recover the system studied by \cite{Grappin_1993_PRL} and \cite{Meyrand_2024_JPP}.
To revert to the Eulerian frame, we simply set the frame velocity to zero ($U_{\mathrm{fr}} = U_{\mathrm{fr},0} = 0$).

However, not all frames are equally relevant.
While flux tube simulations offer the most detailed insights into wave behavior, they are computationally expensive, and simplified models can be more efficient.
Then, to understand how waves launched by the Sun evolve—and to match the steady-state solution that arises in a full flux tube simulation—it is natural to adopt the $z^+$ frame, in which $U_{\rm fr} = U + v_{\rm A}$, and to allow the $z^+$ fluctuations to decay in this frame without forcing~$\boldsymbol{z}^+$. The resulting mean square value of~$z^+$ in the box, $\langle z^+_{\rm box}(t)^2\rangle$, 
can be used to estimate the average mean-square value of $z^+$ in the solar wind,
$\langle z^+_{\rm sw}(r)^2\rangle$,  by setting $\langle z^+_{\rm sw}(r)^2\rangle = \langle z^+_{\rm box}(t(r))^2\rangle$, where $t(r)$ is the time at which the moving box is at heliocentric distance $r$.

We note that, in a given simulation, if the box moves more slowly than $U + v_{\rm A}$, it will take longer to reach a given heliospheric distance, leading to excessive wave cascading and energy dissipation compared to that which could occur in the flux tube.
To correct this, energy would have to be injected into the box. On the other hand, if the frame moves faster than $U + v_{\rm A}$, energy dissipation needs to be artificially increased.
In the special case of a fixed Eulerian frame, the system must be forced so that $z^+_\mathrm{rms}$ fluctuates about a steady mean, as it does at fixed~$r$ in the solar wind.

The Eulerian frame should provide a powerful method for investigating local turbulent processes, allowing us to focus on the system after it has reached a steady state.
In this regime, time evolution becomes irrelevant, and computational resources can be allocated to enhancing resolution and extending the inertial range without the burden of tracking the dynamics of different spatial locations as required in moving box simulations.
It is likely important in such simulations to choose the radial extent of the box to be longer than the distance that $z^-$ fluctuations can propagate during their cascade timescales to prevent the same $z^-$ and $z^+$ fluctuations from interacting mutliple times. This approach should also enable the system to reach a steady state after just a few eddy turnover times, maximizing computational efficiency.

%----------------------------------------------
% NEW SECTION
%----------------------------------------------
\section{Scalar Formulation} \label{sec:Scalar}

When solving the IRMHD equations on a computer, one can reduce the amount of computer time required by first rewriting the equations in terms of scalar potentials. To this end, we define the Elsasser potentials $\zeta^\pm$ via the equation
\begin{equation}
    \label{eq:zetapm}
    \boldsymbol{z}^\pm (\boldsymbol{x}, t) = \frac{\hat{\boldsymbol{e}}_{z} \times \boldsymbol{\nabla} \zeta^\pm (\boldsymbol{x}, t)}{\alpha}.
\end{equation}
Eq. (\ref{eq:zetapm}) implies that the velocity and magnetic-field fluctuations can be expressed as
\begin{equation}
    \delta \boldsymbol{u} (\boldsymbol{x}, t) = \frac{\hat{\boldsymbol{e}}_{z} \times \boldsymbol{\nabla} \Phi (\boldsymbol{x}, t)}{\alpha}, \quad \text{and} \quad \delta \boldsymbol{b} (\boldsymbol{x}, t) = \frac{\hat{\boldsymbol{e}}_{z} \times \boldsymbol{\nabla} \Psi (\boldsymbol{x}, t)}{\alpha},
\end{equation}
where $\Phi$ and $\Psi$ are two stream functions that are determined up through an arbitrary additive constant and that satisfy the equation $\zeta^\pm = \left(\Phi \mp \Psi \right) $.
Similarly, we introduce the wave-action potentials $\zeta_{\rm f}^\pm$, which satisfy the equation
\begin{equation} \label{def:wave-action-potentials}
    \boldsymbol{f}^\pm (\boldsymbol{x}, t) = \hat{\boldsymbol{e}}_{z} \times \boldsymbol{\nabla} \zeta_{\rm f}^\pm (\boldsymbol{x}, t), \quad \text{with} \quad \zeta_{\rm f}^\pm \equiv \frac{1 \pm \eta^{1/2}}{\eta^{1/4}} \zeta^\pm.
\end{equation}
Finally, we define the field-aligned Elsasser and wave-action vorticities, $\Omega^\pm$ and $\Omega_{\rm f}^\pm$, via the equations
\begin{subequations}
\begin{align}
    \Omega^\pm (\boldsymbol{x}, t) &= \hat{\boldsymbol{e}}_{z} \cdot \left( \boldsymbol{\nabla} \times \boldsymbol{z}^\pm \right) = \nabla_\perp^2 \zeta^\pm, \\
    \Omega_{\rm f}^\pm (\boldsymbol{x}, t) &= \hat{\boldsymbol{e}}_{z} \cdot \left( \boldsymbol{\nabla} \times \boldsymbol{f}^\pm \right) = \nabla_\perp^2 \zeta_{\rm f}^\pm,
\end{align}
\end{subequations}
where $\nabla_\perp^2 \equiv \partial_{xx} + \partial_{yy}$.
Throughout this paper, we henceforth assume $\nabla_\perp \gg \partial_z \gg H_i^{-1}$, with $i = \{A, B, \rho\}$.
Given the previous assumptions, taking the curl of Eqs. (\ref{eq:Heinemann-Olbert-Lagrangian}), and keeping only the terms of the lowest order, we obtain the second main result of this paper:
\begin{equation} \label{eq:IRMHD-wave-action-potentials}
    \left[ \frac{\partial}{\partial t} + \left( U - U_\mathrm{fr} \pm v_A \right) \frac{U_{\mathrm{fr},0}}{U_\mathrm{fr}} \frac{\partial}{\partial z_0} \right] \Omega_{\rm f}^\pm - \frac{U \pm v_A}{2 H_A} \Omega_{\rm f}^\mp = \left\{ \partial_j \zeta_{\rm f}^\pm, \partial_j \zeta^\mp \right\} + \left\{ \Omega_{\rm f}^\pm, \zeta^\mp \right\},
\end{equation}
where we have employed the Einstein summation convention with the index $j=\{x,y\}$, and where the Poisson bracket $\left\{f, g\right\} \equiv \hat{\boldsymbol{e}}_{z} \cdot \left(\boldsymbol{\nabla} f \times \boldsymbol{\nabla} g \right)$ for arbitrary functions $f$ and $g$. Eqs. (\ref{eq:IRMHD-wave-action-potentials}) can be divided in three parts:
\begin{enumerate}
    \item
        The squared brackets on the left-hand side correspond to a linear advection of wave-action vorticities $\Omega_{\rm f}^\pm$ at speed $\left( U - U_\mathrm{fr} \pm v_A \right) U_{\mathrm{fr},0} / U_\mathrm{fr} $.
    \item
        The last term on the left-hand side describes non-WKB reflections, which act as a source for inward-propagating waves that are correlated to the original wave.
        This correlation leads to the phenomenon often referred to as anomalous coherence \citep[see, e.g.,][]{Velli_1989_PRL, Verdini_2009_ApJ, Perez_2013_ApJ, Chandran_2019_JPP}.
        This term also disrupts the conservation of energy and cross-helicity for the turbulent fluctuations.
        However, these conservation laws are restored when the energy and cross helicity of the background are included \citep{Chandran_2015_ApJ}.
    \item
        The nonlinear terms capture two key effects.
        First (by order of appearance), the vortex stretching effect, arising from interactions between opposite populations, influences the growth or decay of vorticities \citep{Zhdankin_2016_PoP}.
        As highlighted by \cite{Schekochihin_2022_JPP}, the nonlinear term forces $\Omega_{\rm f}^+$ and $\Omega_{\rm f}^-$ equally but in opposite directions at every point in space and time, resulting in a negative correlation between the two vorticities.
        This interaction systematically promotes current sheets over shear layers.
        Second, nonlinear advection describes how each vorticity is transported by the field of the other.
\end{enumerate}

We note that Eq. (\ref{eq:IRMHD-wave-action-potentials}) becomes problematic near the Alfv\'en critical point~$r_{\rm A}$, where $\Omega_{\rm f}^-$ and~$\zeta_{\rm f}^-$ vanish, and where the value of $\zeta^-$ in the equation for~$\partial \Omega_{\rm f}^+/\partial t$ would need to be computed by dividing~$\zeta_{\rm f}^-$ by $1-\eta^{1/2}$, which also vanishes at~$r=r_{\rm A}$. To avoid the need to evaluate~$\zeta_{\rm f}^-/(1-\eta^{1/2})$ in moving-box simulations that transit past $r=r_{\rm A}$, one could
replace the equation for $\partial \Omega_{\rm f}^-/\partial t$ by an equivalent equation for $\partial \Omega^-/\partial t$
\footnote{
The equivalent equation for the Elsasser vorticity of the `–' wave is expressed as
$ \left[ \partial_t + \left( U - U_\mathrm{fr} - v_{\rm A} \right) \frac{U_{\mathrm{fr},0}}{U_\mathrm{fr}}  \partial_{z_0} \right] \Omega^- + \left(U + v_{\rm A} \right) \left( \frac{\Omega^-}{4 H_\rho} - \frac{\Omega^+}{2 H_{\rm A}}\right) = \left\{ \partial_j \zeta^-, \partial_j \zeta^+ \right\} + \left\{ \Omega^+, \zeta^- \right\}.$
}.

%-----------------------------------------
% NEW SECTION
%-----------------------------------------
\section{Linear system} \label{sec:Linear}

In this subsection, we consider a rectangular box at some initial position in the Sun's frame.
We take the radial extent $\Delta r$ of the box to be much smaller than the distance over which the equilibrium properties change appreciably.
Anticipating future numerical simulations, we make the approximation that the fluctuations in the box satisfy triply periodic boundary conditions.
We further take $H_{\rm A}$ to be a nonzero constant, but we treat $v_{\rm A}$ and $U$ as uniform within the box.
Although a spatially uniform $v_{\rm A}$ is inconsistent with a nonzero $H_{\rm A}$, the variations in $v_{\rm A}$ that we are neglecting are $\sim \Delta r / H_{\rm A} \ll 1$, and, we conjecture, unimportant—similarly to Boussinesq's approximation.
In the general case, simple linear sinusoidal solutions to Eqs. (\ref{eq:IRMHD-wave-action-potentials}) cannot be obtained because $U(z(t))$, and $v_A(z(t))$ depends on time in a non trivial way
\footnote{
However, when all variables depend similarly on $z(t)$, a change of variables can remove the time dependence, allowing for a linear analysis. This applies to the super-Alfvénic solar wind, where $v_A \ll U \sim \text{constant}$ and $v_A$ scales as $1/t$ \citep{Meyrand_2024_JPP}.
}.
We therefore choose to focus this section on the Eulerian frame, in which the box remains at a fixed position in the Sun's frame.
We now consider a small perturbation of the wave-action potentials $\zeta_{\rm f}^\pm = \zeta^\pm_{{\rm f}0} \mathrm{e}^{ \mathrm{i} \left( \boldsymbol{k} \cdot \boldsymbol{x} - \omega t \right)}$.
Upon substituting this perturbation into Eqs. (\ref{eq:IRMHD-wave-action-potentials}), we obtain the matrix equation
\begin{equation}
\begin{pmatrix}
 \mathrm{i} \left( \omega' - \omega_A \right) & (U + v_A)(2H_A)^{-1} \\
 (U - v_A)(2H_A)^{-1} & \mathrm{i}  \left( \omega' + \omega_A \right) \\
\end{pmatrix}
\begin{pmatrix}
 \zeta^+_{{\rm f}0} \\
 \zeta^-_{{\rm f}0} \\
\end{pmatrix} = 0,
\end{equation}
with $\omega' \equiv \omega - k_z U$, and $\omega_A \equiv k_z v_A$. This linear set of equations supports waves forward and backward propagating modes of frequency
\begin{equation} \label{eq:dispersion-relation—Heinemann-Olbert}
    \omega_\pm'(k_z) = \pm v_A \sqrt{ k_z^2 + \frac{1 - \mathcal{M}_A^2}{4 H_A^2 }}.
\end{equation}

This equation generalizes the Alfvén-wave dispersion relation to allow for finite Alfvén-speed scale heights, which makes the waves dispersive.
For nonzero values of $k_\|$, the second term under the square-root sign in Eqs. (\ref{eq:dispersion-relation—Heinemann-Olbert}) is a small correction to the first term as~$\Delta r \ll H_{\rm A}$.
However, when $k_\|=0$, that second term is the only term that survives, leading to purely growing or damped solutions when $M_A > 1$, and to low-frequency oscillatory solutions when $M_A < 1$.

We note that the eigenfunctions that diagonalize the linear system are
\begin{equation}
    \Theta_k^\pm = \mp \frac{U-v_{\rm A}}{2 H_{\rm A}} \xi_k^+ + \mathrm{i} \left( \omega_+' \mp \omega_A \right) \xi_k^-.
\end{equation}
Unlike RMHD, in which the eigenmodes are $\zeta^\pm_k$, here $\Theta_k^\pm$ combine both $\xi^+_{k}$ and $\xi^-_{k}$.
This distinction introduces an important new feature: the governing equations for the eigenfunctions $\Theta_k^\pm$ inherently describe nonlinear interactions between both co-propagating and counter-propagating waves.

%---------------------------------------------------------
% NEW SECTION
%---------------------------------------------------------
\section{Discussion and conclusion} \label{sec:Conclusion}

In this paper, we have developed a family of local approximations to the nonlinear equations derived by \cite{Heinemann_1980_JGR}, including a box that moves radially outward at an arbitrary velocity. The use of Clebsch coordinates ensures independence from radial magnetic field line spreading and greatly simplifies the system's geometry, allowing us to derive Eqs. (\ref{eq:IRMHD-wave-action-potentials}), which are a relevant extension for both sub- and super-Alfvénic plasmas, regardless of whether the turbulence is balanced or imbalanced, and whether the medium is homogeneous or not. 

There are several numerical approaches to solve Eqs. (\ref{eq:IRMHD-wave-action-potentials}), each with its advantages and limitations.
Flux-tube simulations, for example, are the most comprehensive as they cover the full relevant range in $r$, making them ideal for capturing the full complexity of the turbulence \citep[see e.g.][]{Perez_2013_ApJ, van_Ballegooijen_2016_ApJ, van_Ballegooijen_2017_ApJ, Chandran_2019_JPP}.
However, they are incredibly expensive computationally.
This is because they require an extremely large number of radial grid points, and also because Alfvén waves must travel from the Sun out to the Alfvén critical point and back a few times before the turbulence settles down into a statistical steady state \citep{Perez_2013_ApJ}.
The Eulerian box—a small stationary box in the expanding solar wind—will allow a wide inertial range and is particularly useful for studying local turbulence dynamics and phenomena like cascade rates and anomalous coherence. 
The trade-off, however, is that it requires forcing to sustain the root mean square value of $\boldsymbol{z}^+$, denoted $z^+_{\rm rms}$, at a chosen level, and $z^+_{\rm rms}$ and the parallel box length must be large enough that $\boldsymbol{z}^-$ fluctuations to a good approximation cascade and dissipate before crossing the box to avoid spurious interactions with their parent outward-propagating waves.
Additionally, it lacks radial tracking, so while one can get insight into inertial-range dynamics, it misses out on how the turbulence evolves with heliospheric distance and thus cannot predict coronal/solar wind heating rates.
Finally, the moving box method generalizes the well known expanding-box model to a non-constant solar wind's speed, and strikes a balance between these two approaches.
This approach corresponds to the accelerating expanding box model derived by \cite{Tenerani_2017_ApJ} in the specific limit of incompressible and strongly anisotropic fluctuations. However, our model introduces a key advantage: it offers the flexibility to select a reference frame that is best suited to the problem at hand. In contrast, the accelerating expanding box model is inherently tied to the plasma bulk flow, which becomes less useful in the sub-Alfvénic wind, as discussed in Section \ref{subsec:transformation}.
Like the Eulerian box, the moving box method gives a large inertial range and can explore anomalous coherence, but it also captures the decay of turbulence as it propagates away from the Sun, in a way that is independent of the forcing mechanism—if the $z^+$ frame is used.
One of its main strengths is the ability to track turbulence radially as the simulation progresses.
However, it comes with the drawback of limited statistics at any given radius, since the box moves along with the plasma flow.
Each method offers its own unique insights, but the choice ultimately depends on the specific aspect of turbulence one is trying to study.

Finally, our focus has been on Alfvén waves.
To extend this work we could cover the Alfvénic cascade down to electron scales, deriving an extension of FLR-MHD \citep{Schekochihin_2019_JPP, Meyrand_2021} to finite scale heights.

%-----------------------------------------------------------
\appendix
%-----------------------------------------------------------
\section{Moving frame transformation} \label{app:Lagrangian}

To understand the transformation to a frame moving at speed $U_\mathrm{fr}(z)$, we introduce new time and radial variable{s} $t'$ and~$z_0$, with~$t^\prime = t$. The coordinates $x$ and $y$ remain unchanged and, unlike the time variable, are not coupled to the transformation, so we ignore them here for brevity.
We start by defining the radial position $z$ of a point initially located at position $z_0$ at time $t_0'$ as a function of $t'$, written as $z \left(t' \, | \, z_0, t_0' \right) $, assuming the point moves outward at speed~$U_{\mathrm{fr}}(z)$. The evolution of $z$ over time is given by:
\begin{equation}
\frac{\partial}{\partial t'} z \left(t' \, | \, z_0, t_0' \right) = U_\mathrm{fr} \left[ z \left(t' \, | \, z_0, t_0' \right) \right].
\end{equation}
This equation can be integrated to yield
\begin{equation}
\label{eq:integral1}
z \left(t' \, | \, z_0, t_0' \right) - z_0 = \int_0^{t'} U_\mathrm{fr} \left[ z \left(t_1 \, | \, z_0, t_0^\prime \right) \right] \mathrm{d}t_1.
\end{equation}
Likewise, if the initial position of the point at time~$t_0^\prime$ is~$z_0 + \Delta$, where~$\Delta $ is some small radial displacement, then the point's position at time~$t^\prime$ satisfies the equation
\begin{equation}
\label{eq:integral2}
z \left(t' \, | \, z_0 + \Delta, t_0' \right) - z_0 - \Delta = \int_0^{t'} U_\mathrm{fr} \left[ z \left(t_1 \, | \, z_0 + \Delta, t_0^\prime \right) \right] \mathrm{d}t_1.
\end{equation}
We define $\delta t$ such that $z \left(t_0' + \delta t \, | \, z_0, t_0' \right) = z_0 + \Delta.$ That is, $\delta t$ is the amount of time required for a point starting at $z_0$  to reach position~$z_0 + \Delta$. It then follows that
\begin{equation}
\label{eq:commutativity}
z \left(t' + \delta t \, | \, z_0, t_0' \right) 
=
z \left(t' + \delta t \, | \, z_0 + \Delta , t_0' + \delta t\right)
= z \left(t' \, | \, z_0 + \Delta, t_0' \right).
\end{equation}
The first equality in (\ref{eq:commutativity}) results from the fact that the point that starts at $(z_0, t^\prime_0)$ moves through spacetime coordinates $(z_0 + \Delta , t^\prime_0 + \delta t)$ on its way to final position $z \left(t' + \delta t \, | \, z_0, t_0' \right) $. The second equality in (\ref{eq:commutativity}) follows from the fact that $U_{\rm fr}(z)$ depends on~$z$ but not~$t^\prime$. When $\delta t$ and~$\Delta$ are infinitesimal,
\begin{equation}
\delta t = \frac{\Delta}{U_\mathrm{fr}(z_0)}.
\end{equation}
Using (\ref{eq:commutativity}) to rewrite the right-hand side of~(\ref{eq:integral2}) and then subtracting~(\ref{eq:integral1}) from~(\ref{eq:integral2}), we obtain
\begin{equation}
\label{eq:integral3}
    z(t^\prime\, |\, z_0 + \Delta, t_0') - z(t^\prime\,|\,z_0, t_0') - \Delta = \int_{t^\prime}^{t^\prime + \delta t} U_\mathrm{fr} \left[ z \left(t_1 \, | \, z_0, t_0^\prime \right) \right] \mathrm{d}t_1 - \int_0^{\delta t} U_\mathrm{fr} \left[ z \left(t_1 \, | \, z_0, t_0^\prime \right) \right] \mathrm{d}t_1.
\end{equation}
Noting that the final terms on the left- and right-hand sides of (\ref{eq:integral3}) cancel and dividing (\ref{eq:integral3}) by $\Delta$, we arrive at 
\begin{equation}
\frac{z(t^\prime \, | \, z_0 + \Delta, t_0') - z(t^\prime \, | \, z_0, t_0')}{\Delta} = \frac{\delta t}{\Delta} U_\mathrm{fr} \left[ z \left(t^\prime \, | \, z_0, t_0^\prime \right) \right].
\end{equation}
Finally, upon taking the limit~$\Delta \rightarrow 0$, we find that
\begin{equation}
\label{eq:stretch}
\left.\frac{\partial z}{\partial z_0}\right|_{t'} = \frac{U_\mathrm{fr}(z)}{U_\mathrm{fr}(z_0)} = \frac{U_\mathrm{fr}}{U_{\mathrm{fr},0}}.
\end{equation}
Equation (\ref{eq:stretch}) describes how the coordinate system is stretched or compressed by the radial variations in~$U_{\rm fr}(z)$.
The Jacobian matrix associated with this change of variables is
\begin{equation}
\label{eq:Jacobian}
    J \equiv \frac{\partial \left( z, t \right) }{\partial \left( z_0, t' \right)} =
    \begin{pmatrix}
        \left.\frac{\partial z}{\partial z_0}\right|_{t'} & \left.\frac{\partial z}{\partial t'}\right|_{z_0}\\
        \left.\frac{\partial t}{\partial z_0}\right|_{t'} & \left.\frac{\partial t}{\partial t'}\right|_{z_0}
    \end{pmatrix}
    =
    \begin{pmatrix}
        \frac{U_\mathrm{fr}}{U_{\mathrm{fr},0}} & U_\mathrm{fr}\\
        0 & 1
    \end{pmatrix}
\end{equation}
It follows from the chain rule that the Jacobian matrix of the inverse transformation is the inverse of the matrix in (\ref{eq:Jacobian}). Hence,
\begin{equation}
\label{eq:invJacobian}
    J^{-1} = \frac{\partial \left( z_0, t' \right)}{\partial \left( z, t \right) } =
    \begin{pmatrix}
        \left.\frac{\partial z_0}{\partial z}\right|_{t} & \left.\frac{\partial z_0}{\partial t}\right|_{z}\\
        \left.\frac{\partial t'}{\partial z}\right|_{t} & \left.\frac{\partial t'}{\partial t}\right|_{z}
    \end{pmatrix}
    =
    \begin{pmatrix}
        \frac{U_{\mathrm{fr},0}}{U_\mathrm{fr}} & -U_\mathrm{fr}\\
        0 & 1
    \end{pmatrix},
\end{equation}
where the final equality follows from taking the inverse of the $2\times 2$ matrix on the right-hand side of~(\ref{eq:Jacobian}). Equation~(\ref{eq:invJacobian}) provides the partial derivatives that we will need in order to transform~(\ref{eq:Clebsch-Heinemann-Olbert}) to the $(z_0, t^\prime)$ coordinate system.  In particular, (\ref{eq:invJacobian}) and the chain rule enable us to write
\begin{subequations}
\begin{align}
    \frac{\partial \boldsymbol{f}^\pm}{\partial t}
    &= \left.\frac{\partial \boldsymbol{f}^\pm}{\partial t'}\right|_{z_0} - U_{\mathrm{fr},0} \frac{\partial \boldsymbol{f}^\pm}{\partial z_0}, \\
    \frac{\partial \boldsymbol{f}^\pm}{\partial z}
    &= \left.\frac{\partial t'}{\partial z}\right|_{t} \left.\frac{\partial \boldsymbol{f}^\pm}{\partial t'}\right|_{z_0} + \left.\frac{\partial z_0}{\partial z}\right|_{t} \left.\frac{\partial \boldsymbol{f}^\pm}{\partial z_0}\right|_{t'} = \frac{U_{\mathrm{fr},0}}{U_\mathrm{fr}} \frac{\partial \boldsymbol{f}^\pm}{\partial z_0}.
\end{align}
\end{subequations}
Upon substituting these expressions into~(\ref{eq:Clebsch-Heinemann-Olbert}) and omitting the prime symbol on~$t^\prime$ to make the notation more compact, we obtain~(\ref{eq:Heinemann-Olbert-Lagrangian}).

%-----------------------
\bibliographystyle{jpp}
\bibliography{bilbio}
\end{document}